\documentclass{aa}

\def\nat{Nature}
\def\ica{Icarus}
\def\jpc{J.~Phys.\ Chem.\ Ref.\ Data}
\def\grl{Geophys.~Res.~Lett.}
\def\asr{Adv.~Space Res.}
\def\aap{A\&A}
\def\apj{ApJ}
\def\jgr{J.~Geophys.~Res.}
\def\atd{Atomic Data and Nuclear Data Tables}
\def\jes{J.~Electron Spectrosc.\ Relat.\ Phenom.}
\def\pra{Phys.~Rev.~A}
\def\jpb{J.~Phys.~B: At.\ Mol.\ Phys.}
\def\aas{A\&AS}
\def\apa{Applied Physics A}
\def\aps{ApJS}
\def\ast{J.~Atm.\ Solar Terr.\ Phys.}
\def\pss{Planet.~Space~Sci.}
\def\<#1>{\relax}
\usepackage{natbib}
\bibpunct{(}{)}{;}{a}{}{,}
\usepackage{graphicx}
\usepackage{psfig}

\advance\voffset by -1.0cm

\begin{document}

\title{Discovery of X--rays from Venus with Chandra}

\author{K. Dennerl \inst{1}
   \and V. Burwitz \inst{1}
   \and J. Englhauser \inst{1}
   \and C. Lisse \inst{2}
   \and S. Wolk \inst{3}
        }

\institute{Max--Planck--Institut f\"ur extraterrestrische Physik,
           Giessenbachstra{\ss}e, 85748 Garching, Germany
           \and
           University of Maryland,
           Department of Astronomy, College Park, MD 20742
           \and
           Chandra X--Ray Center,
           Harvard--Smithsonian Center for Astrophysics,
           60 Garden Street, Cambridge, MA 02138
            }

\offprints{K. Dennerl, \email{kod@mpe.mpg.de}}

\date{Received 28 September 2001 / Accepted 17 January 2002}

\abstract{%
On January 10 and 13, 2001, \object{Venus} was observed for the first time
with an X--ray astronomy satellite. The observation, performed with the
ACIS--I and LETG\,/\,ACIS--S instruments on Chandra, yielded data of high
spatial, spectral, and temporal resolution. Venus is clearly detected as a
half--lit crescent, with considerable brightening on the sunward limb. The
morphology agrees well with that expected from fluorescent scattering of solar
X--rays in the planetary atmosphere. The radiation is observed at discrete
energies, mainly at the O--K$_\alpha$ energy of 0.53~keV. Fluorescent
radiation is also detected from C--K$_\alpha$ at 0.28~keV and, marginally, from
N--K$_\alpha$ at 0.40~keV. An additional emission line is indicated at
0.29~keV, which might be the signature of the C $1s\to\pi^{\star}$ transition
in CO$_2$ and CO. Evidence for temporal variability of the X--ray flux was
found at the $2.6\,\sigma$ level, with fluctuations by factors of a few times
indicated on time scales of minutes. All these findings are fully consistent
with fluorescent scattering of solar X--rays. No other source of X--ray
emission was detected, in particular none from charge exchange interactions
between highly charged heavy solar wind ions and atmospheric neutrals, the
dominant process for the X--ray emission of comets. This is in agreement with
the sensitivity of the observation.
}

\authorrunning{K. Dennerl et al.}

\maketitle

\section{Introduction}

The first detection of X--ray emission from a planetary atmosphere
came as a surprise, when unexpectedly high background radiation was
observed in 1967 during a daytime stellar X--ray survey by a rocket
\citep{68jgr012}\<Grader..>. This radiation was correctly attributed
to X--ray fluorescence of the Earth's atmosphere.
\cite{70nat068}\<Aikin> estimated that the same process should also
cause other planetary atmospheres to glow in X--rays, although the
expected flux at Earth orbit would be extremely small and only
detectable with sophisticated instrumentation. X--ray fluorescence of
the \object{Earth} atmosphere, however, became a well--known component of
the X--ray background of satellites in low--Earth orbit. Its properties
and its impact on observations were studied in detail by
\cite{88aap286}\<Fink..> and \cite{93apj361}\<Snowden..>.\\[2.2ex]
The detection of unexpectedly bright X--ray emission from comets
\citep{96sci001,97sci002,97apj353}\<Lisse.. Dennerl.. Mumma..>
has led to increased interest in X--ray studies of solar system
objects. With its carbon and oxygen rich atmosphere, the absence of a
strong magnetic field, and its proximity to the Sun, Venus represents
a close planetary analog to a comet. Dissociative recombination of
O$_2^+$ in the Venus ionosphere leads to a hot oxygen exosphere out to
over 4000~km \citep{85asr007}\<Russel..>, resembling a cometary coma.
To investigate the X--ray properties of Venus, we performed a
pioneering observation with the X--ray astronomy satellite Chandra.\\[2.2ex]
Orbiting the Sun at heliocentric distances of 0.718\,--\,0.728
astronomical units (AU), the angular separation of Venus from the Sun,
as seen from Earth, never exceeds 47.8 degrees. While the observing
window of imaging X--ray astronomy satellites is usually restricted to
solar elongations of at least $60\degr$, Chandra is the first such
satellite which is able to observe as close as $45\degr$ from the
limb of the Sun. Thus, with Chandra an observation of Venus with an
imaging X--ray astronomy satellite became possible for the first time.
The observation was scheduled to take place around the time of
greatest eastern elongation, when Venus was $47\degr$ away from the
Sun. At that time it appeared optically as a very bright (-4.4~mag),
approximately half--illuminated crescent with a diameter of $23\arcsec$
(Table~\ref{obscxo}).

\begin{table*}
\caption[]{Journal of observations and observing geometry}
\label{obscxo}
\begin{tabular}{cccccccccc}
\noalign{\vspace*{-1ex}}
\hline
\noalign{\vspace*{0.5ex}}
obsid & date & time & exp time & instrument &
$r$ & $\Delta$ & phase & elong & diam \\
& 2001 & [UT] & [s] & & [AU] & [AU] & [$^{\circ}$] & [$^{\circ}$] & [$\arcsec$] \\
\noalign{\vspace*{0.5ex}}
\hline
\noalign{\vspace*{1.0ex}}
2411 & Jan\,10 & 19:32:47\,--\,21:11:55 & \phantom{0}5\,948 &
LETG\,/\,ACIS--S & 0.722 & 0.734 & 85.0 & 47.0 & 22.7
\\
2414 & Jan\,10 & 21:24:30\,--\,23:00:26 & \phantom{0}5\,756 &
LETG\,/\,ACIS--S & 0.722 & 0.734 & 85.0 & 47.0 & 22.8
\\
\phantom{0}583 & Jan\,13 & 12:39:51\,--\,15:57:40 & 11\,869 &
ACIS--I & 0.721 & 0.714 & 86.5 & 47.1 & 23.4 \\
\noalign{\smallskip}
\hline
\end{tabular} \\[1.2ex]
obsid: Chandra observation identifier, exp time: exposure time,
$r$: distance from Sun, $\Delta$: distance from Earth,\\
phase: angle Sun--Venus--Earth, elong: angle Sun--Earth--Venus,
diam: angular diameter
\end{table*}

\section{Observation and data analysis}

Venus is the celestial object with the highest optical surface
brightness after the Sun and a highly challenging target
for an X--ray observation with Chandra, as the X--ray detectors there
(CCDs and microchannel plates) are also sensitive to optical light.
Suppression of optical light is achieved by optical blocking filters
which, however, must not attenuate the X--rays significantly. The
observation was originally planned to use the back--illuminated
ACIS--S3 CCD, which has the highest sensitivity to soft (E$<$1.4~keV)
X--rays, for direct imaging of Venus, utilizing the intrinsic energy
resolution for obtaining spectral information. Before the
observation was scheduled, however, it turned out that the optical
filter on this CCD would not be sufficient for blocking the extremely
high optical flux from Venus. Therefore, half of the observation (obsid 583,
cf.\,Table~\ref{obscxo}) was performed with the front--illuminated CCDs of the
ACIS--I array (I1 and I3), which are less sensitive to soft X--rays, but which
are also significantly less affected by optical light contamination.

In order to avoid any ambiguity in the X--ray spectrum due to residual
optical light, the other
half of the observation (obsid 2411 and 2414, cf.\,Table~\ref{obscxo})
was performed with the low
energy transmission grating (LETG), where the optical and X--ray
spectra are completely separated by dispersion. To minimize the risk
of telemetry overload, the area around the zero order image
(contaminated by optical light) was not transmitted to Earth. The
telescope was offset by $3\farcm25$ from its nominal aimpoint, to
shift the most promising spectral region around the dispersed 0.53~keV
O--K$_\alpha$ emission line well into the back--illuminated CCD S3
(Fig.\,\ref{acis2}). The combination of direct imaging and
spectroscopy with the transmission grating made it possible to obtain
complementary spatial and spectral information within the available
total observing time of 6.5 hours.

At the time of the observation, Venus was moving across the sky
with a proper motion of $2\farcm6$/hour. As the CCDs were read out
every 3.2~s, there was no need for continuous tracking. The spacecraft
was oriented such that Venus would move parallel to one side of the
CCDs and perpendicular to the dispersion direction in the LETG
observation. To keep Venus well inside the $8\farcm3$ field of view (FOV)
of ACIS--S perpendicular to the dispersion direction, Chandra was
repointed at the middle of the LETG\,/\,ACIS--S observation
(Table~\ref{obscxo}). For ACIS--I with its larger $16\farcm9\times16\farcm9$
FOV, no repointing during the observation was necessary.
As the photons were recorded time--tagged, an individual post--facto
transformation into the rest frame of Venus is possible. This was done
with the geocentric ephemeris of Venus as computed with the JPL
ephemeris calculator.\footnote{available at
\tt http://ssd.jpl.nasa.gov/cgi-bin/eph }
Correction for the parallax of Chandra was done with the orbit
ephemeris of the delivered data set.
For the whole analysis we used events with Chandra standard grades and
excluded bright pixels. The fact that all observations were performed
with CCDs with intrinsic energy resolution made it possible to suppress
the background with high efficiency.

\begin{figure}
\centerline{\resizebox{1.0\hsize}{!}{\psfig{file=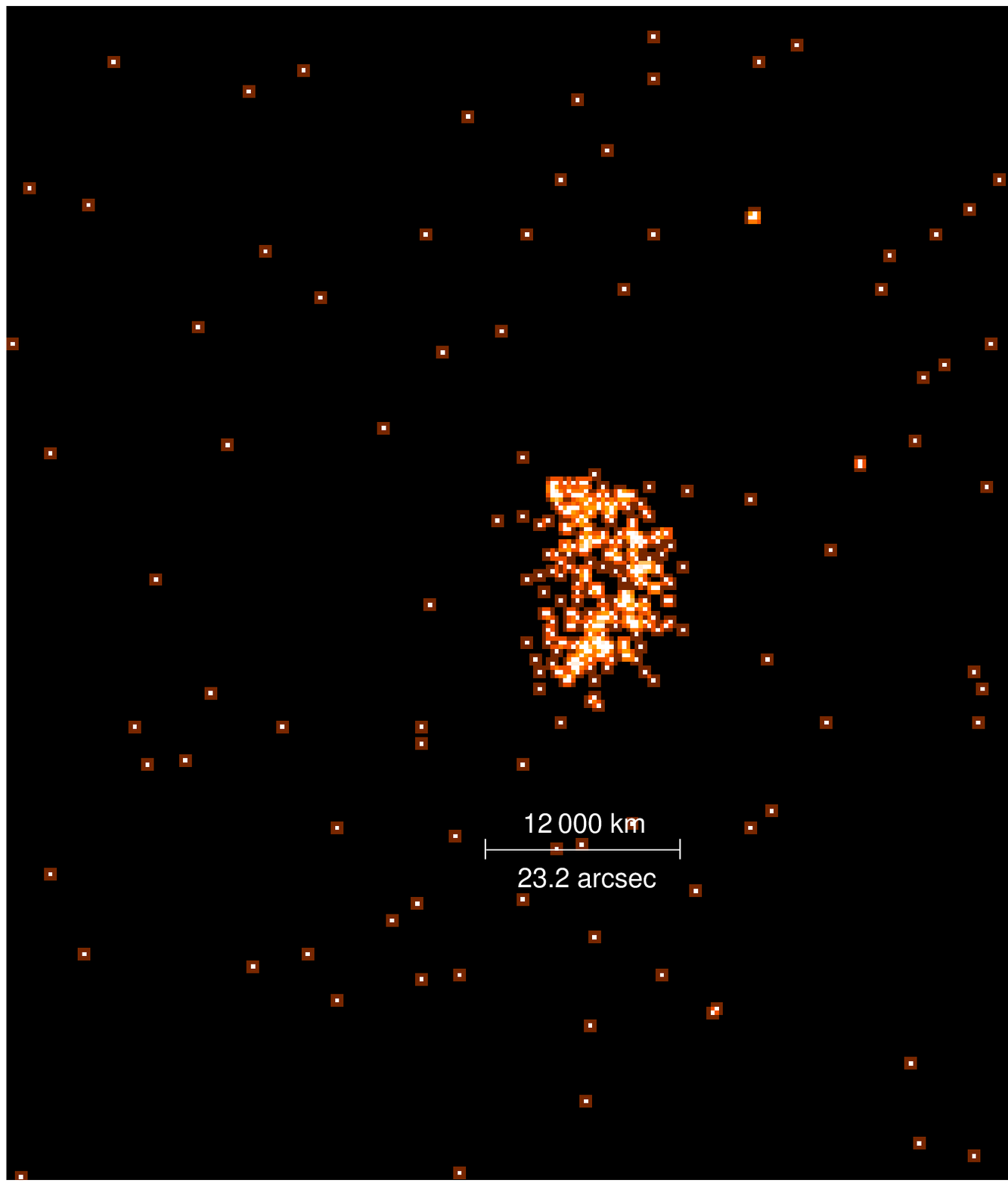,clip=}}}
\caption{First X--ray image of Venus, obtained with Chandra
\mbox{ACIS--I} on 13 January 2001.
}
\label{v00z}
\end{figure}

\begin{figure}
\hbox to \hsize{\hfil\vbox{\hsize=0.85\hsize
\resizebox{\hsize}{!}{\psfig{file=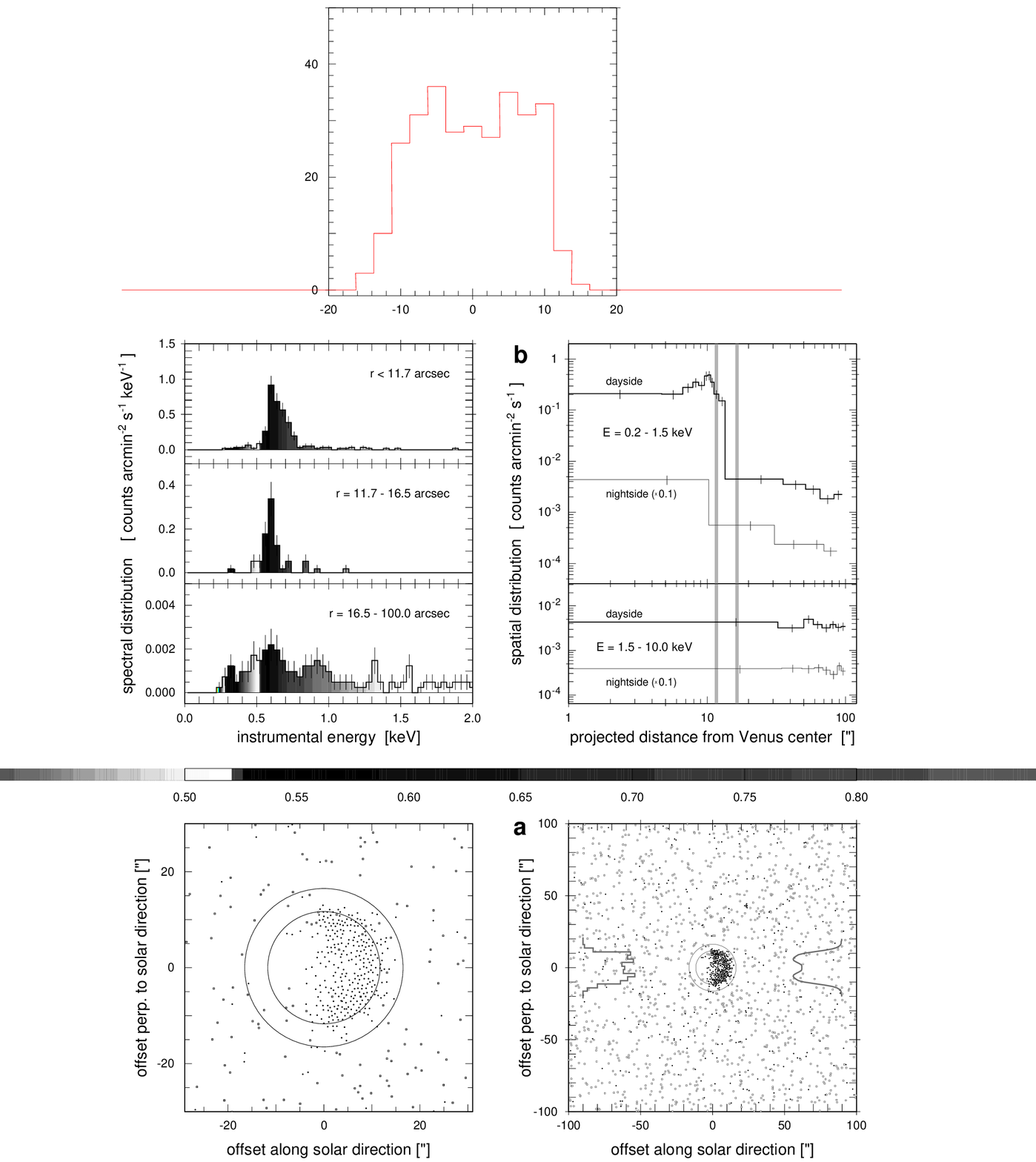,%
bbllx=295pt,bblly=72pt,bburx=515pt,bbury=285pt,clip=}} \\
\resizebox{\hsize}{!}{\psfig{file=MS1955f2.eps,%
bbllx=295pt,bblly=315pt,bburx=515pt,bbury=566pt,clip=}}
}\hfil}
\caption{Spatial distribution of photons around Venus in the ACIS--I
observation. {\bf a)} All photons in the energy range
0.2\,--\,10.0~keV; those with $E\le1.5\mbox{ keV}$
are marked with black dots, while photons with $E>1.5\mbox{ keV}$ are
plotted as open squares.
In some cases the symbols have been
slightly shifted (by typically less than $1\arcsec$) to minimize overlaps.
Two circles are shown, the inner one, with $r=11\farcs7$, indicating the
geometric size of Venus, and the outer one, with $r=16\farcs5$, enclosing
practically all photons detected from Venus.
The histogram to the left shows the distribution of photons with
$E\le1.5\mbox{ keV}$ from the outer circle, projected onto the y axis, with
$2\farcs5$ resolution; the smoothed version to the right was used as a template
for fitting the LETG spectrum.
{\bf b)} Spatial distribution of the photons in the soft
($E=0.2$\,--\,1.5~keV) and hard ($E=1.5$\,--\,10.0~keV) energy range,
in terms of surface brightness along radial rings around Venus,
separately for the `dayside' (offset along solar direction $>0$) and
the `nightside' (offset $<0$). For better clarity the nightside
histograms were shifted by one decade downward. The bin size was
adaptively determined so that each bin contains at least 24 counts.
Thick vertical lines mark the radii of $11\farcs7$ and $16\farcs5$ of the
circles in (a).}
\label{vn4spa}
\end{figure}

\section{Results}

\subsection{Morphology}

In the X--ray image (Fig.\,\ref{v00z}) the crescent of Venus is clearly
resolved and allows detailed comparisons with the optical appearance.
An optical image (Fig.\,\ref{simsum}\,e), taken at the same phase
angle, shows a sphere which is slightly more than half illuminated,
closely resembling the geometric illumination, with the brightness
maximum well inside the crescent. In the X--ray image the sphere
appears to be less than half illuminated. The most striking
difference, however, is the pronounced limb brightening, which is
particularly obvious in the surface brightness profiles
(Fig.\,\ref{vn4spa}\,b) and in the smoothed X--ray image
(Fig.\,\ref{simsum}\,d). For a quantitative understanding of this
brightening we performed a simulation of the appearance of Venus in
soft X--rays, based on fluorescent scattering of solar X--ray
radiation. This simulation will be described in Sect.\,4.

The ACIS--I observation is well suited to investigate the X--ray
environment of Venus outside the bright crescent, an area unaffected
by optical loading. In Fig.\,\ref{vn4spa} the distribution of photons
within a $200\arcsec\times200\arcsec$ region around Venus is shown, together
with the evolution of the surface brightness with increasing distance
from Venus, for a soft ($E=0.2$\,--\,1.5~keV) and a hard
($E=1.5$\,--\,10.0~keV) band. In the soft band there is some
indication that the surface brightness at $r=16\farcs5$\,--\,$100\arcsec$
decreases with increasing distance from Venus, both at the day- and
nightside. The fact that the corresponding distributions in the hard
band are flat argues against the possibility that this drop is caused
by inhomogeneities in the overall sensitivity. If the drop is not
caused by other instrumental effects, e.g., the PSF of the telescope,
then this could be evidence for an extended X--ray halo around Venus.
This will be discussed further in Sect.\,5.

\subsection{Spectrum}

The ACIS--I data clearly show that the X--ray spectrum of Venus is very
soft: images accumulated from photons with energies $E>1.5\mbox{ keV}$
show no enhancement at the position of Venus
(cf.\,Fig.\,\ref{vn4spa}\,b). We determine a
$3\,\sigma$ upper limit of $2.5\cdot10^{-4}\mbox{ counts/s}$ to any
flux from Venus in the 1.5\,--\,2.0~keV energy range. The corresponding
value for $E=2.0$\,--\,8.0~keV is $5.6\cdot10^{-4}\mbox{ counts/s}$.
Further spectral analysis of the ACIS--I data is complicated by the
presence of optical loading.%
\footnote{Although ACIS--I is equipped with an efficient filter for
blocking optical light, the optical brightness of Venus (surface
brightness: 1.5~mag per square arcsecond) is so extreme that a small
fraction of optical photons succeed in penetrating the filter. These
photons deposit charge in the CCD in addition to the X--ray photons,
causing a systematic increase of the apparent energy. As a
consequence, the spectrum of the photons within the optical extent of
Venus ($r<11\farcs7$) extends to higher energies than the one accumulated
near the outer boundary ($11\farcs7<r<16\farcs5$), where only a fraction of
photons (projected there by the PSF of the telescope) contributes to
optical loading. Despite this contamination, both spectra indicate the
presence of a bright emission line near 0.6~keV.\vspace*{2ex}}

\begin{figure*}
\vspace*{1mm}
\resizebox{\hsize}{!}{\psfig{file=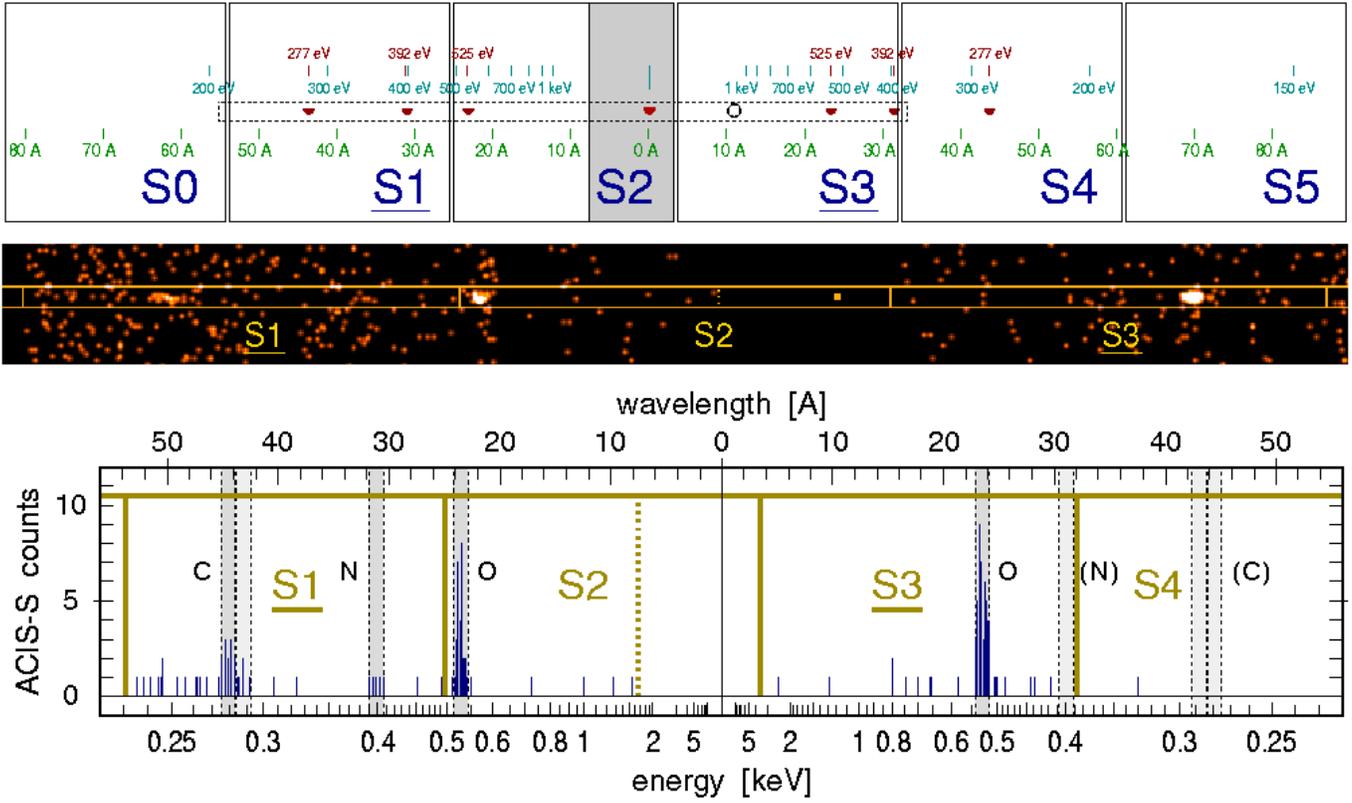}}
\vspace*{-2mm}
\caption{%
{\bf a)} Expected LETG spectrum of Venus on the ACIS--S array. S1 and
S3 are back--illuminated CCDs with increased sensitivity at low
energies (underlined), while the others are front--illuminated. The
nominal aimpoint, in S3, is marked with a circle. The aimpoint was
shifted by $3.25'$ into S2, to get more of the fluorescent lines
covered by back--illuminated CCDs. During the two parts of the
observation, Venus was moving perpendicularly to the dispersion
direction. In order to avoid saturation of telemetry, the shaded area
around the zero order image in S2 was not transmitted. Energy and
wavelength scales are given along the dispersion direction. Images of
Venus are drawn at the position of the C, N, and O fluorescence lines, with
the correct size and orientation. The dashed rectangle indicates the
section of the observed spectrum shown below.
{\bf b)} Observed LETG spectrum of Venus, smoothed with a Gaussian
function with $\sigma=2\farcs0$. The two bright crescents symmetric to
the center are images in the line of the O-K$_{\alpha}$ fluorescent
emission, while the elongated enhancement at left is at the position
of the C-K$_{\alpha}$ fluorescent emission line. The Sun is at bottom.
Vertical lines mark the extent of the individual CCDs. The position of
the zero order image (not transmitted) is indicated by a small square
in S2. Contamination from higher orders is negligible.
{\bf c)} Spectral scan along the region outlined above. Scales
are given in keV and \AA. The observed C, N, and O fluorescent emission
lines, at the energies listed in Tab.\,\ref{phflx},
are enclosed by dashed lines; the width
of these intervals matches the size of the Venus crescent
($22\farcs8$). The thick lines mark the borders of the individual CCDs.
}
\label{acis2}
\end{figure*}

Anticipating this possibility, and in order to fully utilize the
spectroscopic capabilities of Chandra, we performed also spectroscopic
observations with the LETG. Fig.\,\ref{acis2}\,a shows what we
expected to see in the dispersed first order spectra: images of the
Venus crescent in the C--K$_\alpha$, N--K$_\alpha$ and O--K$_\alpha$
emission lines (the energies were taken from \cite{67rmp001}\<Bearden>).
Fig.\,\ref{acis2}\,b shows the observed spectrum. The
photons from both LETG observations were transformed into the rest
frame of Venus and superimposed. Bright pixels and bright columns were
excluded, and the intrinsic energy resolution of ACIS--S was used to
suppress the background.%
\footnote{This was done by selecting only events within
$[E_{\rm min},E_{\rm max}]$, with the energy boundaries depending on
the displacement $x$ from the zero order position:

\hspace*{2ex}$E_{\rm min} = 270\mbox{ eV} + (E-277\mbox{ eV}) \cdot 0.8065$\\
\hspace*{2ex}$E_{\rm max} = 350\mbox{ eV} + (E-277\mbox{ eV}) \cdot 1.0887$

\hspace*{2ex}$E = 1.239\cdot10^3\mbox{ eV}/\lambda[{\rm nm}]
\,,\;\,
\lambda/[{\rm nm}] = 5.602\cdot10^{-3}\cdot\left|x[\arcsec]\right|$
}
\noindent
This spectrum, which is completely uncontaminated by optical light, clearly
shows that most of the flux comes from O--K$_\alpha$ fluorescence. As this
flux is monochromatic, images of the Venus crescent (illuminated from bottom)
show up along the dispersion direction. The intensity profile along this
direction is shown in Fig.\,\ref{acis2}\,c.

\begin{table*}
\caption[]{Number of photons, derived fluxes, and line energies}
\label{phflx}
\vspace*{-2ex}
\begin{tabular}{ccccccccc}
\noalign{\smallskip}
\hline
\noalign{\vspace*{0.5ex}}
grating & line & CCD & type & eff area & photons & photon flux & energy flux &
energy \\
\noalign{\vspace*{0.3ex}}
\hline
\noalign{\vspace*{1ex}}
LETG & C--K$_{\alpha}^{\,\star}$ & S1 & BI & \phantom{0}14 cm$^2$ &
       $18\pm8$ & $1.1\pm0.5$ & $4.8\pm2.1$ &
       $278.6\,^{+1.3}_{-1.1}\mbox{ eV}$ \\[0.6ex]
LETG & N--K$_{\alpha}$ & S1 & BI & \phantom{00}5 cm$^2$ &
       $\phantom{0}4\pm2$ & $0.6\pm0.3$ & $4.0\pm2.0$ &
       $398.5\,^{+4.1}_{-5.9}\mbox{ eV}$ \\[0.6ex]
LETG & O--K$_{\alpha}$ & S2 & FI & \phantom{00}7 cm$^2$ &
       $32\pm7$ & $3.8\pm0.8$ & $32.\pm7.0$ &
       $527.7\,^{+4.6}_{-3.8}\mbox{ eV}$ \\[0.6ex]
LETG & O--K$_{\alpha}$ & S3 & BI & \phantom{0}21 cm$^2$ &
       $50\pm9$ & $2.1\pm0.4$ & $17.2\pm3.1$ &
       $527.2\,^{+1.3}_{-1.4}\mbox{ eV}$ \\[0.6ex]
none & (O--K$_{\alpha}$) & I\,1,3 & FI & 135 cm$^2$ &
       $274\pm17$ & $1.7\pm0.1$ & $14.4\pm0.9$ & --- \\
\noalign{\smallskip}
\hline
\end{tabular} \\[1.5ex]
LETG: spectroscopy with grating, first order;
none: direct imaging (not simultaneous with spectroscopy);\\
BI: back--illuminated, FI: front--illuminated;
eff area: effective area used for the calculation;\\
photon flux in units of $10^{-4}$ ph cm$^{-2}$ s$^{-1}$;
energy flux in units of $10^{-14}$ erg cm$^{-2}$ s$^{-1}$\\
$^{\star}$ There is some evidence for a secondary emission line
at $288\pm4\mbox{ eV}$; the photons from this line\\
were included in the flux determination for C--K$_{\alpha}$. \\[1.5ex]

\end{table*}

For the determination of the line energies we proceeded as follows: from the
ACIS--I observation, we accumulated all photons with $E\le1.5\mbox{ keV}$
within a circle of $r=16\farcs5$ around Venus along the solar direction, to get
the intensity profile perpendicular to the solar direction
(Fig.\,\ref{vn4spa}a), i.e., along the dispersion axis in the LETG observation
(Fig.\,\ref{acis2}). This profile, which shows a characteristic central dip
due to the effect of limb brightening, was then smoothed with a cubic spline
function and used as a template for the spectral fit. For each emission line,
the position and the normalization of the template were determined as free
parameters by $\chi^2$ minimization. The dispersion was expressed in
arcseconds, and the width of the template was reduced by 3\% (to take the
change of the apparent size into account; Tab.\,\ref{obscxo}) and kept fixed.
Errors were determined by increasing the minimum $\chi^2$ by 1.0, with the
normalization as free parameter. The results, converted into energies, are
listed in Tab.\,\ref{phflx}.

The observed line energies are higher than the values of
\cite{67rmp001}\<Bearden>, but in all cases the deviation is less than
$1.5\,\sigma$. A better agreement, with deviations of less than $1.0\,\sigma$,
is achieved with \cite{70nat068}\<Aikin>, who quoted emission energies of 278.4,
392.9, and 526.0~eV for C, N, and O. Recent determinations of the dominant
emission line of atomic oxygen (for the 1s2s$^2$2p$^5$($^3$P$^{\mbox{o}}$)
configuration) yielded energies in the range 526.8\,--\,528.3~eV \citep[and
references therein]{98jpb001}\<McLaughlin and Kirby>, which is in excellent
agreement with the $527.2^{+1.3}_{-1.4}\mbox{ eV}$ found in the LETG\,--\,S3
spectrum of Venus. If the oxygen atom is embedded in a molecule, this energy
is slightly shifted: to
527.3~eV for CO$_2$ \citep{97apa001}\<Nordgren>,
526.8~eV for O$_2$  \citep{98jpb001}\<McLaughlin>, and
525.0\,--\,526.0~eV for CO \citep{97phr006}\<Skytt..>.
These shifts are too small to be discriminated with
the currently available spectral resolution, which is mainly limited by the
statistical uncertainty due to the low number of detected photons. The
situation is similar for carbon and nitrogen, where the following values were
recently found:
279.2~eV for CO \citep{97phr006}\<Skytt..> and
393\,--\,394~eV for N$_2$ \citep{97apa001}\<Nordgren>.

Emission line spectra of molecules composed of C, N, and O atoms are
characterized by
an additional fluorescence line, caused by the $1s\to\pi^{\star}$
transition following core excitation. In CO, this line is much more pronounced
at the C than at the O fluorescence energy
\cite[e.g.][Figs. 5 and 6]{97phr006}\<Skytt..>.
The energies for the $1s\to\pi^{\star}$ transition are
287.4~eV for C in CO \citep{84jes001}\<Sodhi and Brion>,
290.7~eV for C in CO$_2$ \citep{87jes002}\<Hitchcock and Ishi>,
401.1~eV for N$_2$, 534.2~eV for O in CO \citep{84jes001}\<Sodhi and Brion>, and
535.4~eV for O in CO$_2$ \citep{87phr005}\<McLaren..>.
We do not see such an additional line at N and O. The image of Venus at
the energy of carbon, however, appears elongated along the dispersion direction
(Fig.\,\ref{acis2}b), and the spectrum (Fig.\,\ref{acis2}c) does
indicate the presence of a secondary emission line at $288\pm4\mbox{ eV}$,
consistent with the energy of the C $1s\to\pi^{\star}$ transition in
CO$_2$ and CO.

Table~\ref{phflx} summarizes the number of detected
photons in the individual lines together with the derived
fluxes. The specified uncertainties are
$1\,\sigma$ errors from photon statistics; they do not include
uncertainties in the effective areas at $E<0.6\mbox{ keV}$.
This could be the reason for the $1.9\,\sigma$
deviation between the two O--K$_\alpha$ fluxes observed with S2 and
S3, because both parts of the dispersed spectrum were recorded
simultaneously. For C--K$_\alpha$ the number of recorded photons is
sufficient for detection, but not high enough for a reliable flux
determination. For N--K$_\alpha$ a marginal detection is only
possible because the few photons were recorded exactly at the expected
position. The fact that no N--K$_\alpha$ photons are detected at the
corresponding mirror site may be related to inhomogeneities in the
ACIS--S low energy response, in particular close to the CCD borders.
The last row in Table~\ref{phflx} contains the values for the
direct imaging observation with ACIS--I. Despite the lower
energy resolution and the problem with optical loading, it is very
likely that most of the flux came from O--K$_\alpha$. The difference
between this flux and the (non--simultaneously obtained) O--K$_\alpha$
fluxes from the LETG observations may be related to the X--ray
variability of Venus (Sect.\,3.3).

It is interesting to compare the total X--ray flux from Venus with
the optical flux. The visual magnitude $-4.4\mbox{ mag}$
corresponds to an optical flux of
$f_{\rm opt}=1.5\cdot10^{-3}\mbox{ erg cm}^{-2}\mbox{ s}^{-1}$.
Adopting a total X--ray flux of
$f_{\rm x}=3.0\cdot10^{-13}\mbox{ erg cm}^{-2}\mbox{ s}^{-1}$,
we get
$$ f_{\rm x}/f_{\rm opt} = L_{\rm x}/L_{\rm opt} = 2\cdot10^{-10} $$

\noindent
Taking into account that the energy of a K$_{\alpha}$ photon exceeds
that of an optical photon by two orders of magnitude, we find that on
average there is only one X--ray photon among $5\cdot10^{11}$ photons
from Venus. This extremely small fraction of X--ray versus optical
flux, combined with the soft X--ray spectrum and the proximity of
Venus to the Sun, illustrates the challenge of observing Venus in
X--rays. The X--ray flux is emitted in just three narrow emission
lines. Outside these lines the $L_{\rm x}/L_{\rm opt}$ ratio is even
orders of magnitude lower.

\begin{figure}
\centerline{\resizebox{\hsize}{!}{\psfig{file=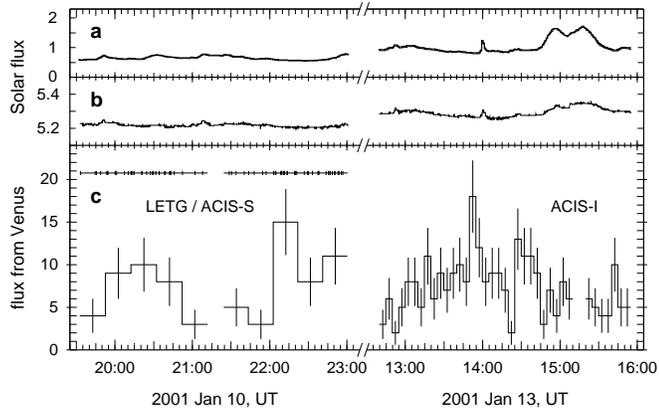,clip=}}}
\caption{Temporal behaviour of the soft X--ray flux from the Sun and
Venus. {\bf a)} 1\,--\,8~\AA\ (1.55\,--\,12.4~keV) solar flux in
$10^{-3}$ erg cm$^{-2}$ s$^{-1}$ at 1.0~AU, as measured with GOES--8
and GOES--10. {\bf b)} 1\,--\,500~\AA\ (0.025\,--\,12.4~keV)
solar flux in $10^{10}$ photons cm$^{-2}$ s$^{-1}$ scaled to 1.0~AU,
as measured with SOHO/SEM. The times in (a) and (b)
were shifted by $+240$~s and $+230$~s for Jan 10 and 13, to take the
light travel time delay between Sun\,$\to$\,Venus\,$\to$\,Chandra and
Sun\,$\to$\,SOHO into account.
{\bf c)} X--ray flux from Venus as observed with Chandra
LETG\,/\,ACIS--S and ACIS--I. The ACIS--S data are shown with 1189.6~s
and 1151.2~s time resolution for the first and second part (to avoid
partially exposed time bins). Only photons at the energy of the
O\,--\,$K_{\alpha}$ emission line were taken from the first order LETG
spectra; the background contribution is negligible
(cf.\,Fig.\,\ref{acis2}). Individual photon arrival times are
indicated at top together with the exposure duration. The ACIS-I count
rates, shown with 300~s time resolution, were derived by extracting
all photons below 1~keV from a circle of $16\farcs5$ radius centered at
Venus. The interruption at 15:15~UT is caused by Venus crossing the
gap between CCD I1 and I3. With less than 0.1 background events per
time bin the background is negligible.
}
\label{vlc}
\end{figure}

\subsection{Temporal variability}

While there is practically no variation of the optical flux from Venus
on time scales of hours and less, the X--ray flux shows indications
for pronounced variability on time scales of minutes
(Fig.\,\ref{vlc}\,c). A Kolmogorov--Smirnov test yields probabilities
of only 1\% for both the observation with LETG\,/\,ACIS--S and ACIS--I
that the count rates are just statistical fluctuations around a
constant value. As variability of the 1\,--\,10~keV solar flux by a
factor of two on time scales of minutes is not uncommon
(e.g.\,Fig.\,\ref{vlc}\,a), we expect the scattered solar X--rays from
Venus to exhibit a similar variability. However, a direct comparison
with the solar flux monitored simultaneously with GOES--8 and GOES--10
(Fig.\,\ref{vlc}\,a) and SOHO/SEM (Fig.\,\ref{vlc}\,b) does not show
an obvious correlation. This may be related to the fact that solar
X--rays are predominantly emitted from localized regions and that
Venus saw a solar hemisphere which was rotated by $48\fdg0$
(LETG\,/\,ACIS--S) and $46\fdg5$ (ACIS--I) from the solar hemisphere facing
Earth.
Differences from the broad band solar X--ray flux (as measured with
GOES/SOHO) may also arise due to the fact that the X--ray flux from Venus
responds very sensitively to variability of the solar flux in a narrow
spectral range just above the K edges. As will be shown in the next section,
this is particularly the case for O--K$_{\alpha}$: while
the C--K$_{\alpha}$ emission increases by only 7\% if the coronal temperature
rises by 8\%, the O--K$_{\alpha}$ emission increases by 33\%.

\section{Modeling the X--ray appearance of Venus}

Estimates on the X--ray luminosity of Venus due to scattering and
fluorescence of solar x--rays have recently been made by
\cite{01grl006}\<Cravens and Maurellis>. We are, however, not
aware of detailed predictions of how Venus would appear
in X--rays. For comparison with the observed image we performed
a numerical simulation of fluorescent scattering of solar X--rays
in the atmosphere of Venus. We did not consider elastic scattering,
as the corresponding luminosity is one order of magnitude below
the fluorescence luminosity, according to
\cite{01grl006}\<Cravens and Maurellis>, and in agreement with
the observed LETG spectrum (Fig.\,\ref{acis2}\,c).

The ingredients to the model are the composition and density structure
of the Venus atmosphere, the photoabsorption cross sections and
fluorescence efficiencies of the major atmospheric constituents, and
the incident solar spectrum.

\subsection{Venus atmosphere}

We adopted the Venus model atmosphere from \cite{83venus001}\<Seiff>,
where the density in the lower and middle atmosphere, i.e., between
the surface and a height of 100~km, is tabulated in steps of 2~km for
different latitudes, while for the upper atmosphere, between 100~km
and 180~km, it is tabulated in steps of 4~km for two solar zenith
angles sza (subsolar: $\mbox{sza}<50^{\circ}$ and antisolar:
$\mbox{sza}>120^{\circ}$). For
$50^{\circ}\leq\mbox{sza}\leq120^{\circ}$ we interpolated the density
exponentially. In order to calculate the number density of C, N, and O atoms,
we used the following values for the composition of the atmosphere:
65.2\% oxygen, 32.6\% carbon and 2.2\% nitrogen. As the main
constituents, C and O, are contained in CO$_2$, we assumed this
composition to be homogeneous throughout the atmosphere.

\begin{figure}
\resizebox{\hsize}{!}{\psfig{file=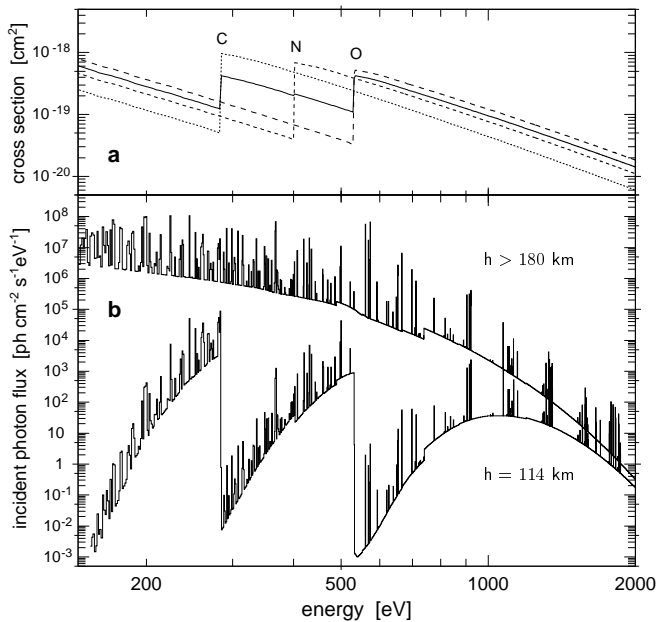,clip=}}
\caption{{\bf a)} Photoabsorption cross sections $\sigma_{\rm C}$,
$\sigma_{\rm N}$, $\sigma_{\rm O}$ for C, N, and O (dashed lines),
and $\sigma_{\rm CNO}$ for the chemical composition of the Venus
atmosphere (solid line).
{\bf b)} Incident solar X--ray photon flux on top of the Venus
atmosphere ($h>180\mbox{ km}$) and at 114~km height (along subsolar
direction; below). The spectrum is plotted in 1~eV bins. At 114~km, it
is considerably attenuated just above the K$_{\alpha}$ absorption
edges, recovering towards higher energies.
}
\label{crscflx}
\end{figure}

\subsection{Photoabsorption cross sections}

\begin{figure}
\resizebox{\hsize}{!}{\psfig{file=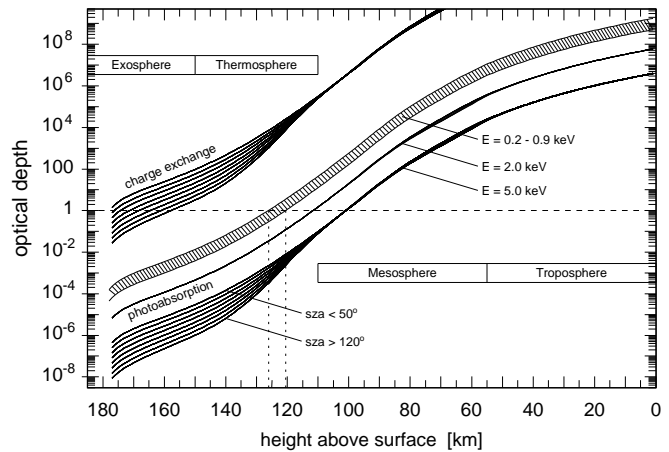,clip=}}
\caption{Optical depth
$\tau=\tau_{\rm C}+\tau_{\rm N}+\tau_{\rm O}$
of the Venus model atmosphere with respect to charge exchange (above)
and photoabsorption (below), as seen from outside. The upper/lower
boundaries of the hatched area refer to energies just above/below the
C and O edges (cf.\ Fig.\,\ref{crscflx}\,a). For better clarity the
dependence on the solar zenith angle (sza) is only shown for
$E=5.0\mbox{ keV}$; the curves for the other energies refer to
$\mbox{sza}<50^{\circ}$. The dashed line, at $\tau=1$, marks the
transition between the transparent ($\tau<1$) and opaque ($\tau>1$)
range. For a specific energy, the optical depth increases by at least
12 orders of magnitude between 180~km and the surface. For charge
exchange interactions a constant cross section of
$3\cdot10^{-15}\mbox{ cm}^2$ was assumed.
}
\label{atmo2}
\end{figure}

The values for the photoabsorption cross sections were taken from
\cite{79apj362}\<Reilman and Manson>.
We supplemented them with data from \cite{95jpc002}\<Chantler> at
energies close to the K edges. From these values and the C, N, and O contributions
listed above, we computed the effective photoabsorption cross section of the
Venus atmosphere (Fig.\,\ref{crscflx}\,a). This, together with the atmospheric
density structure, yielded the optical depth of the Venus atmosphere, as seen
from outside (Fig.\,\ref{atmo2}).

There is quite some discrepancy in the literature about the K--shell binding
energies in C, N, and O atoms. These energies are affected by the outer electrons and
thus depend on whether the element is in an atomic, molecular, or solid state.
The values 283.8, 401.6, 532.0~eV for C, N, O, which \cite{95jpc002}\<Chantler>
computed for isolated atoms,
are in good agreement with the values 283.84, 400, and 531.7~eV
found by \cite{82atd001}\<Henke> and used by, e.g.,
\cite{83apj193}\<Morrison and McCammon>.
However, \cite{91apj366}\<Gould and Jung> found significantly higher
K--threshold energies for isolated C, N, and O atoms: 297.37~eV, 412.36~eV,
and 546.02~eV. According to \cite{93apj361}\<Snowden and Freyberg>,
the values 400~eV and 532~eV of \cite{82atd001}\<Henke> refer to
molecular nitrogen and oxygen, while \cite{91phr007}\<Ma..> quote an
ionization potential of 409.938~eV for N$_2$ (and 296.080~eV for C in CO).
A compilation by \cite{79atd002}\<Sevier>
lists calculated values of 296.94~eV for atomic carbon,
410.7~eV and 411.88~eV for atomic nitrogen, 411.2~eV for N$_2$,
545.37~eV for atomic oxygen, and 542.2~eV for O$_2$.
An accurate treatment of the K--edge is further complicated by the presence of
considerable fine--structure: detailed calculations of the inner--shell
photoabsorption of oxygen by \cite{98jpb001}\<McLaughlin and Kirby> show that
already the atomic state contains an impressive amount of resonance structure
around the K--edge.

In order to estimate the consequences of all these uncertainties, we ran our
simulation also with the following, alternative set of K--edge energies:
$ E_{K_{\rm C}} = 296.1\mbox{ eV}, E_{K_{\rm N}} = 409.9\mbox{ eV},
  E_{K_{\rm O}} = 544.0\mbox{ eV}. $
In both cases we assumed that the energy will be emitted at
279.2, 393.5 and 527.3~eV, according to recent determinations and in
agreement with the observed LETG spectrum (Sect.\,3.2).
We found no significant differences in the results (Tab.\,\ref{flxs}).

\subsection{Solar radiation}

The solar spectra for 2001 January 10 and 13 were derived from
SOLAR\,2000
\citep{00ast001}\<Tobiska..>.\footnote{available at {\tt http://SpaceWx.com/}}
These spectra, best estimate daily average values, do not
differ much between both dates. To improve the coverage towards
energies $E>100\mbox{ eV}$, we computed synthetic spectra with
the model of \cite{85aap287}\<Mewe..> and aligned them with the
SOLAR\,2000 spectra in the range 50\,--\,500~eV, by adjusting the
temperature and intensity. We derived a coronal temperature of
$120\pm10\mbox{ eV}$. The spectrum, scaled to the heliocentric
distance of Venus, is shown in Fig.\,\ref{crscflx}\,b (upper curve),
with a bin size of 1~eV, which we used in order to preserve the
spectral details.

\subsection{Model grid}

The high dynamic range in the optical depth of the Venus atmosphere
(Fig.\,\ref{atmo2}) requires a model with high spatial resolution. For
the simulation we choose a right--handed coordinate system $(x,y,z)$
with the center of Venus at (0,0,0) and the Sun at $(0,0,+\infty)$. We
sample the irradiated part of the Venus atmosphere with a grid of
volume elements of size $s_x=s_y=s_z\equiv s_g=1\mbox{ km}$.

Fig.\,\ref{atmo2} shows that the atmosphere becomes optically thick for
X--rays with $E<5\mbox{ keV}$ already at heights above 100~km. This
means that solar X--rays do not reach the atmosphere below 100~km,
where the density shows some latitudinal dependence. Above 100~km,
however, the density of the model atmosphere depends only on the height
and the solar zenith angle. This simplifies the calculation: instead
of computing the solar irradiation of the volume elements
$V(x_i,y_j,z_k)$, it is only necessary to do this for volume elements
$V(r_{ij},z_k)$, with $r_{ij}=(x_i^2+y_j^2)^{1/2}$.

The whole information about the irradiation of the atmosphere can thus
be computed and stored in volume elements $V_{ik}=V(r_i,z_{k_i})$,
$i=1..n$ and $k_i=1..m_i$, with
$$n={\rm int}[R_{\rm v}/s_g]\quad\mbox{for}\quad r_i \le R_{\rm v}
\quad\mbox{and} $$
$$m_i=\left\{
  \begin{array}{l}
    {\rm int}\left[{1\over s_g}
     \left(\left(R_{\rm v}^2-r_i^2\right)^{1/2}\!
     -\left(r_{\rm v}^2-r_i^2\right)^{1/2}\right)\right]
     \mbox{ for } r_i \le r_{\rm v} $$ \\[2ex]
    {\rm int}\left[{2\over s_g}\left(
     R_{\rm v}^2-r_i^2\right)^{1/2}\right]
     \quad\qquad\qquad\mbox{for } r_{\rm v} < r_i \le R_{\rm v} $$
  \end{array} 
\right. $$
\noindent
Here $r_{\rm v}=6052.0\mbox{ km}$ is the radius of Venus,
$h_{\rm v}=180\mbox{ km}$ is the height of the model atmosphere,
and $R_{\rm v} = r_{\rm v} + h_{\rm v}$.

\smallskip
With this grid the calculation is performed in two steps: in the first
step the solar radiation absorbed in each volume element is calculated
and stored. In the second step an image of Venus is accumulated for a
particular phase angle by integrating the emission and subsequent
absorption of the corresponding volume elements along the line of
sight.

\subsection{Simulation}

\smallskip
The first step is performed by propagating the irradiation for each
column $i$\/ from the top of the atmosphere along the $-z$ direction,
i.e., away from the Sun. For the center of the corresponding volume
element $V_{ik}$, the mass densitity is calculated from the height
above the surface and the solar zenith angle, by exponential
interpolation of the nearest tabulated grid points in the Venus model
atmosphere, and converted into a number density $n_{ik}$ of the sum of
C, N, and O atoms. From the column densities
$$ \rho_{\ell,ik} = n_{ik} \cdot s_g \cdot f_{\ell}, \quad
\mbox{with } \ell=\rm{C},\rm{N},\rm{O} $$
and
$ f_{\rm C} = 0.326$, $f_{\rm N} = 0.022$, $f_{\rm O} = 0.652 $,
the optical depths $\tau_{\ell,ik}(E)$ are computed 
$$ \tau_{\ell,ik}(E) = \rho_{\ell,ik} \cdot \sigma_{\ell}(E)
   \;,\quad
   \tau_{ik}(E) = \sum_{\ell=1}^3 \tau_{\ell,ik}(E) $$
\noindent
to derive the attenuated solar spectrum $f^{\rm sun}_{ik}(E)$\,:
$$ f^{\rm sun}_{ik}(E) = \left\{
\begin{array}{ll}
   f^{\rm Sun}(E) & \hfill\mbox{for}\quad k=m_{i} \\[0.4ex]
   f^{\rm sun}_{i,k+1}(E) \,e^{-\tau_{i,k+1}(E)} &
                          \quad\mbox{for}\quad 0<k<m_{i}
\end{array} \right.
$$
\noindent
and the flux $f^{\rm abs}_{\ell,ik}(E)$ of absorbed photons which
interact with K$_{\ell}$ electrons:
$$ f^{\rm abs}_{\ell,ik}(E) =
   f^{\rm sun}_{ik}(E) \cdot 
\left\{\begin{array}{cl}
       0 & \quad\mbox{for}\quad E < E_{K_{\ell}} \\
       1-e^{-\tau_{\ell,ik}(E)}&\quad\mbox{for}\quad E\ge E_{K_{\ell}}
       \end{array} \right. $$

\noindent
Here $f^{\rm Sun}(E)$ is the solar spectrum scaled to the
heliocentric distance of Venus (Fig.\,\ref{crscflx}\,b), and
$E_{K_{\ell}}$ are the energies of the K absorption edges
\citep{95jpc002}\<Chantler>:\,%
\footnote{See Sect.\,4.2 for an alternative set of $E_{K_{\ell}}$.}
$$ E_{K_{\rm C}} = 283.8\mbox{ eV},\quad
   E_{K_{\rm N}} = 401.6\mbox{ eV},\quad
   E_{K_{\rm O}} = 532.0\mbox{ eV}. $$

\begin{figure}
\resizebox{\hsize}{!}{\psfig{file=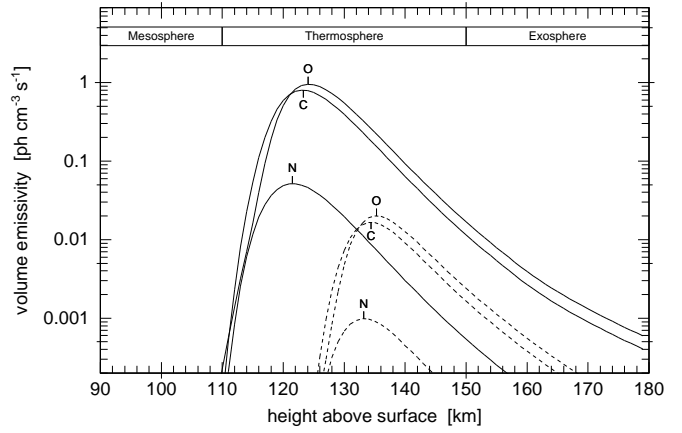,clip=}}
\caption{Volume emissivities of C, N, and O K$_{\alpha}$ fluorescent photons
at zenith angles of zero (subsolar, solid lines) and $90\degr$
(terminator, dashed lines) for the incident solar spectrum of
Fig.\,\ref{crscflx}\,b. The height of maximum emissivity rises with
increasing solar zenith angles because of increased path length and
absorption along oblique solar incidence angles. In all cases maximum
emissivity occurs in the thermosphere, where the optical depth depends
also on the solar zenith angle (Fig.\,\ref{atmo2}).
}
\label{volem}
\end{figure}

\noindent
Only a fraction of the photons which interact with K$_{\ell}$
electrons cause subsequent K$_{\ell}$ shell fluorescence emission:
$$ f^{\rm em}_{\ell,ik}(E) =
   y_{\ell} \cdot f^{\rm abs}_{\ell,ik}(E) $$
with the fluorescent yields $y_{\rm C} = 0.0025$, $y_{\rm N} = 0.0055$
and $y_{\rm O} = 0.0085$ \citep{79jpc001}\<Krause>. The resulting volume
emissivity of fluorescence photons is shown in Fig.\,\ref{volem} for the
subsolar atmospheric column and for a column at the terminator.

As the fluorescence photons are emitted at an energy $E_{\ell}$ which
is below the corresponding K edge $E_{K_{\ell}}$
(see Sect.\,3.2 and 4.2), they are
not subject to subsequent K shell absorption by the same element. K
shell absorption by lighter elements, however, is possible. All
photons are subject to elastic scattering, but as this process is
nearly isotropic, only weakly energy dependent and characterized
by cross sections which are several orders of magnitude smaller than
for photoabsorption, it should not much affect the distribution of
photons along the line of sight. We treat the attenuation of the
reemitted photon flux due to subsequent photoabsorption along the line
of sight in a similar way as we did for the attenuation of the
incident solar flux, but this time only at the three discrete energies
$E_{\ell}$. By sampling the radiation in the volume elements along the
line of sight, starting from the volume element which is farthest away
from the observer, we can then accumulate images of Venus in the three
energies $E_{\ell}$ in, e.g., orthographic projection, for any phase
angle.

\begin{figure*}
\vspace*{1mm}
\centerline{\resizebox{\hsize}{!}{\psfig{file=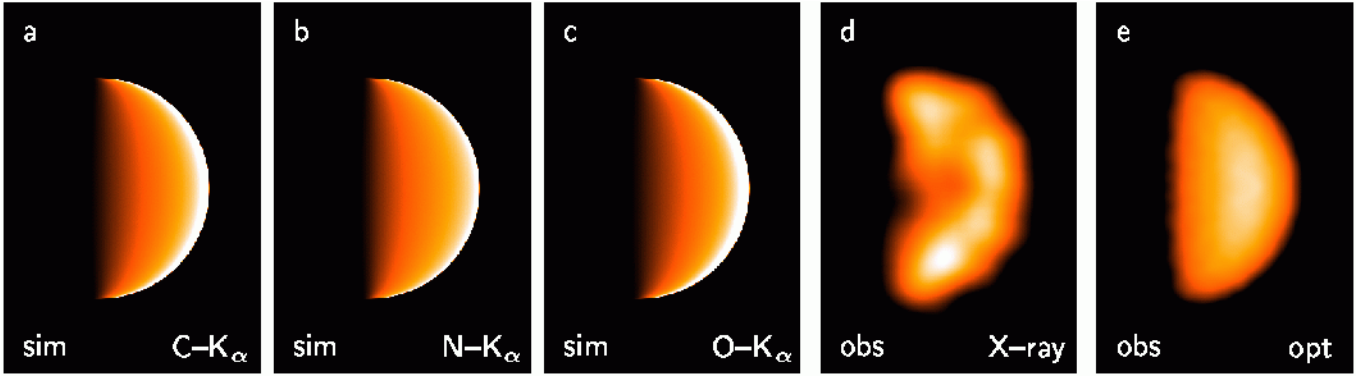,clip=}}}
\caption{{\bf a\,--\,c)} Simulated X--ray images of Venus at
C--K$_{\alpha}$, N--K$_{\alpha}$, and O--K$_{\alpha}$, for
a phase angle $\varphi=86.5^{\circ}$.
The X--ray flux is coded in a linear scale,
extending from zero (black) to
$1.2\cdot10^6\mbox{ ph cm}^{-2}\mbox{ s}^{-1}$ (a),
$5.2\cdot10^4\mbox{ ph cm}^{-2}\mbox{ s}^{-1}$ (b), and
$1.6\cdot10^6\mbox{ ph cm}^{-2}\mbox{ s}^{-1}$ (c),
(white). All images show considerable limb brightening,
especially at C--K$_{\alpha}$ and O--K$_{\alpha}$.
{\bf d)} Observed X--ray image: same as Fig.\,\ref{v00z},
but smoothed with a Gaussian filter with $\sigma=1\farcs8$ and
displayed in the same scale as the simulated images. This image
is dominated by O--K$_{\alpha}$ fluorescence photons.
{\bf e)}~Optical image of Venus, taken by one of the authors (KD)
with a $4''$ Newton reflector on 2001 Jan 12.72 UT, 20 hours before
the ACIS--I observation (cf.\,Table~\ref{obscxo}).
}
\label{simsum}
\end{figure*}

\subsection{Results of the simulation}

The simulated images of Venus at the K$_{\alpha}$ fluorescence
energies C, N, and O are shown in Fig.\,\ref{simsum}\,a\,--\,c.
They agree well with the observed X--ray image (Fig.\,\ref{simsum}\,d),
while the optical image (Fig.\,\ref{simsum}\,e) is characterized by
a different brightness distribution. In X--rays, Venus exhibits
significant brightening at the sunward limb, accompanied by reduced
brightness at the terminator, which causes it to appear less than half
illuminated. This is a consequence of the fact that the volume
emissivity extends into the tenuous, optically thin parts of the
thermosphere and exosphere (Fig.\,\ref{volem}). From there, the volume
emissivities are accumulated along the line of sight without
considerable absorption, so that the observed brightness is mainly
determined by the extent of the atmospheric column along the line of
sight. Detailed comparison of the images shows that the amount of limb
brightening is different for the three energies. This can be
understood in the following way.

If the incident solar spectrum consisted only of photons above the
O--K$_{\alpha}$ edge, then the peak of volume emissivity would occur
at the same height for all fluorescent lines, and this height would be
determined by the spectral hardness of the incident solar flux.
Differences in the height of the volume emissivity peak between C, N,
and O occur due to the presence of photons with energies between the
individual K$_{\alpha}$ edges. Photons with energies between
N--K$_{\alpha}$ and O--K$_{\alpha}$, for example, influence the
atmospheric height of maximum N--K$_{\alpha}$ emission, but do not
affect the O--K$_{\alpha}$ peak. Due to the presence of such photons
in the incident solar spectrum (Fig.\,\ref{crscflx}\,b), which are
affected by less photoabsorption (Fig.\,\ref{crscflx}\,a) and
penetrate deeper into the atmosphere, the nitrogen volume emissivity
peak occurs at the lowest atmospheric heights (Fig.\,\ref{volem}). In
a similar way, the carbon emissivity peak lies just below that of
oxygen.

Although the difference in the atmospheric heights of the individual
peaks is only a few kilometers, this has consequences for the
appearance of Venus in the individual fluorescence lines. At heights
of $\sim 125\mbox{ km}$, the density doubles every 3~km with
decreasing height. The deeper in the atmosphere the emission occurs,
the more absorbing layers are above. This effect is particularly
important at the limb, where the column of absorbing material along
the line of sight reaches a maximum, thus reducing the amount of limb
brightening. Another factor which determines the amount of limb
brightening is the photoabsorption cross section at the fluorescence
energy. This energy is just below the corresponding K--edge
(see Sect.\,3.2 and 4.2).
Fig.\,\ref{crscflx}\,a shows that for the chemical composition of the
Venus atmosphere the photoabsorption cross section for nitrogen
K$_{\alpha}$ fluorescence photons is
about twice as large as that for carbon and oxygen
K$_{\alpha}$ fluorescence photons, causing an additional attenuation
of the limb brightness in the nitrogen image.

The simulations show that the limb brightening depends sensitively on
the density and chemical composition of the Venus atmosphere. Thus,
precise measurements of this brightening will provide direct
information about the atmospheric stucture in the thermosphere and
exosphere. With ACIS--I a brightening of $2.4\pm0.6$ was observed
(Fig.\,\ref{vn4spa}\,b). From the computed images, smoothed with a
Gaussian function with $\sigma=1\farcs0$, we determine corresponding limb
brightenings of 2.0 for C--K$_{\alpha}$, 1.7 for N--K$_{\alpha}$,
and 2.2 for O--K$_{\alpha}$.
The simulated images can also be used to derive the flux from the
whole visible side of Venus in the three energies. Table~\ref{flxs}
shows that the derived flux values highly depend on the coronal
temperature, in particular for O--K$_{\alpha}$. We derive from the
simulation a total flux of $\left(3.6\pm0.9\right)\cdot10^{-13}
\mbox{ erg cm}^{-2}\mbox{ s}^{-1}$ for all three lines.
The corresponding value obtained from the LETG\,/\,ACIS--S observation
(Tab.~\ref{phflx}) is
$\left(4.1\pm0.8\right)\cdot10^{-13}\mbox{ erg cm}^{-2}\mbox{ s}^{-1}$ or
$\left(2.6\pm0.4\right)\cdot10^{-13}\mbox{ erg cm}^{-2}\mbox{ s}^{-1}$,
depending on whether we take the O--K$_{\alpha}$ flux from the S2 or
S3 CCD. Considering all the uncertainties, these values are in good agreement
with each other.

\begin{table}
\caption[]{Numerical results of the simulation}
\vspace{-2.3ex}
obtained for the geometry and mean solar activity during the LETG\,/\,ACIS--S
observation, for which a coronal temperature $kT=0.12\pm0.01\mbox{ keV}$ was
derived; the errors are the result of the uncertainty in this temperature. For
the calculation of the energy flux and luminosity, fluorescence line energies
of 279.2, 393.5, and 527.3~eV were used (Sect.\,3.2)\\[1ex]
\label{flxs}
\begin{tabular}{cccccc}
\hline
\noalign{\vspace*{0.5ex}}
line & photon flux & energy flux & luminosity \\
\hline
\noalign{\smallskip}
C--K$_{\alpha}$ & $2.30\pm0.16$ & $10.3\pm0.7$ & $15.6\pm\phantom{0}1.1$~MW \\
N--K$_{\alpha}$ & $0.10\pm0.02$ &$\phantom{0}0.6\pm0.1$ &
$\phantom{0}1.0\pm\phantom{0}0.2$~MW \\
O--K$_{\alpha}$ & $3.10\pm1.03$ & $26.2\pm8.7$ & $39.6\pm13.2$~MW \\
\noalign{\smallskip}
\hline
\noalign{\vspace*{1ex}}
\multispan{4}{with alternative K--edge energies (Sect.\,4.2):\hfil} \\
\noalign{\smallskip}
C--K$_{\alpha}$ & $2.00\pm0.14$ & $\phantom{0}9.0\pm0.6$ &
$13.6\pm\phantom{0}0.9$~MW \\
N--K$_{\alpha}$ & $0.09\pm0.02$ & $\phantom{0}0.6\pm0.1$ &
$\phantom{0}0.9\pm\phantom{0}0.2$~MW \\
O--K$_{\alpha}$ & $3.08\pm1.03$ & $26.0\pm8.7$ & $39.4\pm13.1$~MW \\
\hline
\end{tabular} \\[1.0ex]
photon flux in units of $10^{-4}$ ph cm$^{-2}$ s$^{-1}$\\
energy flux in units of $10^{-14}$ erg cm$^{-2}$ s$^{-1}$
\end{table}

\begin{figure}
\resizebox{\hsize}{!}{\psfig{file=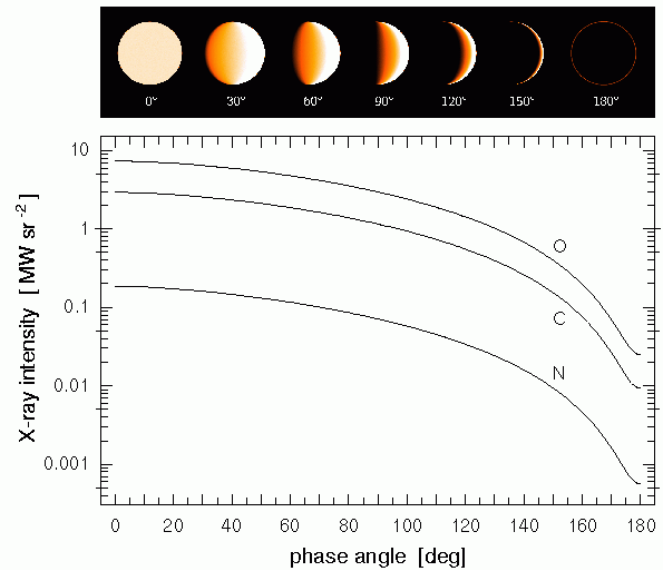,clip=}}
\caption{X--ray intensity of Venus as a function of phase angle, in the
fluorescence lines of C, N, and O. The images at top, all displayed in the
same intensity coding, illustrate the appearence of Venus at O--K$_{\alpha}$
for selected phase angles.}
\label{vnlum1}
\end{figure}

In order to study the angular distribution of the scattered photons, we
computed X--ray images of Venus for different phase angles $\varphi$
(Fig.\,\ref{vnlum1}\,a). We scanned the full range of $\varphi$ from $0\degr$
to $180\degr$ with a step size of $1\degr$, and determined the corresponding
X--ray intensities for the three emission lines by integrating the observed
flux from the images. Fig.\,\ref{vnlum1}\,b shows the result. It is evident
that the intensity declines first very slowly, staying above half of its
maximum value for $\varphi<75\degr$. At $\varphi=90\degr$, the intensity has
dropped to $\sim40\%$. This illustrates that the solar X--rays are preferentially
scattered back towards the Sun. For larger phase angles the decline becomes
faster. Between $170\degr$ and $175\degr$, the intensity drops by a factor of
two, and the X--ray crescent starts to evolve into a thin ring
around the dark planet, which is seen fully developed at $\varphi=180\degr$.
This ring might be observable with sufficiently sensitive solar X--ray
observatories immediately before and after the upcoming Venus transits on 8
June 2004 and 5/6 June 2012. However, such observations would be extremely
challenging, as the intensity of the ring will be only 0.3\% of the fully
illuminated Venus.

By spherically integrating the X--ray intensity for the three energies
(Fig.\,\ref{vnlum1}\,a) over phase angle, we determined the luminosities
listed in Tab.\,\ref{flxs}. The total X--ray luminosity of Venus,
55$\pm$14~MW, agrees well with the prediction of \cite{01grl006}\<Cravens
and Maurellis>, who estimated a luminosity of 35~MW with an uncertainty factor
of about two.

\section{Discussion}

The Chandra data are fully consistent with fluorescent scattering of
solar X--rays in the Venus atmosphere. This is an especially
interesting result when compared with the X--ray emission of comets,
where the dominant process for the X--ray emission is charge exchange
between highly charged heavy ions in the solar wind and cometary
neutrals. The LETG\,/\,ACIS--S spectrum (Fig.\,\ref{acis2})
definitively rules out that a similar process dominates the X--ray flux
from the atmosphere of Venus at heights below $\sim200\mbox{ km}$.

The LETG\,/\,ACIS--S spectrum, however, does not exclude charge
exchange interactions in the outer exosphere of Venus, as they would be
too faint to be detected in the dispersed spectrum. A more sensitive
method for finding charge exchange signatures there is to look for
enhancements of the surface brightness in the environment of Venus. In
fact, the ACIS--I data do show indications for a decrease of the
surface brightness with increasing distance from Venus from $20\arcsec$ to
$100\arcsec$, in the energy range 0.2\,--\,1.5~keV (Fig.\,\ref{vn4spa}). We
find the brightness at $r=20\arcsec$ to exceed that at $r=75\arcsec$ by
$\left(2.7\pm1.0\right)\cdot10^{-3}\mbox{ counts arcmin}^{-2} \mbox{
s}^{-1}$ on the dayside and by
$\left(3.8\pm1.2\right)\cdot10^{-3}\mbox{ counts arcmin}^{-2} \mbox{
s}^{-1}$ on the nightside. Both values are consistent with each other
and yield a mean excess of $\left(3.3\pm0.8\right)\cdot10^{-3}\mbox{
counts arcmin}^{-2} \mbox{ s}^{-1}$.

Observations of comets in the ROSAT all--sky survey 1990\,--\,1991,
also performed at solar maximum, show that the peak surface X--ray
brightness which can be reached by charge exchange is
$\sim 1.0\cdot10^{-13}\mbox{ erg cm}^{-2}\mbox{ s}^{-1} \mbox{
arcmin}^{-2}$ at $r=1\mbox{ AU}$ for an average composition and
density of the low--latitude solar wind \citep{97sci002}\<Dennerl ..>;
it scales with $r^{-2}$. This result is in good agreement with the
theoretical estimate by \cite{97grl001}\<Cravens>. Charge exchange
produces a spectrum consisting of many narrow emission lines. The
overall properties, however, can be approximated by 0.2~keV thermal
bremsstrahlung emission \citep{98pss001}\<Wegmann ..>. By applying
this approximation to the ACIS--I observation, we obtain a maximum
count\-rate due to charge exchange of
$\sim4.6\cdot10^{-3}\mbox{ counts arcmin}^{-2}\mbox{ s}^{-1}$ at
$r=0.72\mbox{ AU}$.

This maximum value is only slightly (by $48\%\pm36\%$) larger than the
excess observed at $r=20\arcsec$, which implies that the exosphere
should be almost collisionally thick 4\,300~km above the surface.
At this height, however, both the hydrogen and the hot oxygen densities
are $\sim10^2\mbox{ cm}^{-3}$
\citep{82ica013,88grl004}\<Bertaux.., Nagy and Cravens>
and thus orders of magnitude too low. Furthermore, the ACIS--I
spectrum of all events within $16\farcs5$ and $100\arcsec$ radius around Venus
(not affected by optical loading) shows no evidence for the spectral
signatures observed in the ACIS--S spectrum of \object{Comet C/1999 S4
(LINEAR)}, which were attributed to charge exchange interactions
\citep{01sci008}\<Lisse..>. We conclude that the observed excess in
the surface brightness is either spurious or produced by other effects.

At heights of 155\,--\,180~km, however, the exosphere of Venus does
become collisionally thick due to the large cross section of charge
transfer interactions (Fig.\,\ref{atmo2}). This implies that if the
flux of highly charged heavy solar wind ions reached these atmospheric
layers, we would indeed observe the maximum flux estimated above. But
even then not more than about 3 photons would have been detected due to
charge exchange from the area of the crescent during the ACIS--I
exposure. Taking the presence of an ionosphere into account, which
shields the lower parts from the solar wind, then even this estimate
appears to be too high. By integrating the X--ray production rate over
altitude, starting at 500~km, the approximate location of the
ionopause, where the density is dominated by the hot oxygen corona,
\cite{00asr006}\<Cravens> estimated a luminosity of $10^5\mbox{ W}$
for the total X--ray luminosity of Venus due to charge exchange. With
the 0.2~keV thermal bremsstrahlung approximation this would result in
a total ACIS--I countrate of $1.6\cdot10^{-5}\mbox{ counts s}^{-1}$,
or only 0.2~counts accumulated during the whole observation. We
conclude that the observation was not sensitive enough for detecting
charge exchange signatures.

It is interesting to compare the X--ray properties of Venus with those
of comets, where the X--ray emission is dominated by the charge
exchange process, while fluorescent scattering of solar X--rays is
negligible. This opposite behaviour is a direct consequence of the
different cross sections of both processes and the way the gas is
distributed.

The cross sections for charge exchange typically exceed
$10^{-15}\mbox{ cm}^2$ and are thus at least three orders of magnitude
greater than for fluorescent emission, which are $10^{-18}\mbox{ cm}^2$
and less in the energy range of interest (Fig.\,\ref{crscflx}\,a). The
gas in a cometary coma is distributed over a much larger volume and
solid angle than in a planetary atmosphere. The particle density in a
coma is too low to reach a column density for efficiently scattering
solar X--rays, but high enough to provide a sufficient number of target
electrons for charge exchange.
The atmosphere of Venus, on the other hand, is so dense that it is
optically thick even to fluorescent scattering (Fig.\,\ref{atmo2}). As
the solar wind ions become discharged already in the outermost parts,
only a tiny fraction of the atmospheric electrons can participate in
the charge exchange process. Additionally, the flux of incident solar
wind ions is reduced by the presence of an ionosphere. Even in the
absence of an ionosphere, the peak X--ray surface flux due to charge
exchange would not exceed that of a comet (with a sufficiently dense
coma), when exposed to the same solar wind conditions. But while the
X--ray bright area is confined to less than one arcmin in diameter in
the case of Venus, the bright part of the cometary X--ray emission can
extend over tens of arcminutes, thus increasing the total amount of
charge exchange induced X--ray photons by two orders of magnitude or
more.

\section{Summary and conclusions}

The Chandra observation clearly shows that Venus is an X--ray source. From the
X--ray spectrum and morphology we conclude that fluorescent scattering of
solar X--rays is the main process for this radiation, which is
dominated by the K$_{\alpha}$ emission lines from C, N, and O, plus some
possible contribution from the C $1s\to\pi^{\star}$ transition in CO$_2$ and
CO. By modeling the X--ray appearance of Venus due to fluorescence, we have
demonstrated that the amount of limb brightening depends sensitively on the
properties of the Venus atmosphere at heights above 110~km. Thus, information
about the chemical composition and density structure of the Venus thermosphere
and exosphere can be obtained by measuring the X--ray brightness distribution
across the planet at the individual K$_{\alpha}$ fluorescence lines. This opens
the possibility of using X--ray observations for remotely monitoring the
properties of regions in the Venus atmosphere which are difficult to
investigate otherwise, and their response to solar activity.

\begin{acknowledgements}
SOLAR\,2000 Research Grade v1.15 historical irradiances are provided
courtesy of W.~Kent Tobiska and SpaceWx.com. These historical
irradiances have been developed with funding from the NASA UARS,
TIMED, and SOHO missions. The SOHO CELIAS/SEM data were provided by
the USC Space Sciences Center. SOHO is a joint European Space Agency,
United States National Aeronautics and Space Administration mission.
\end{acknowledgements}


\end{document}